# Stereoscopic Analysis of the 19 May 2007 Erupting Filament

P. C. Liewer[1] · E. M. De Jong[1] · J. R. Hall[1] · R. A. Howard [2] · W. T. Thompson[3] · J. L. Culhane[4] ·Laura Bone[4] and L. van Driel-Gesztelyi[;4,5,6]

**Abstract**  A filament eruption, accompanied by a B9.5 flare, coronal dimming and an EUV wave, was observed by the *Solar TERrestrial Relations Observatory* (STEREO) on 19 May 2007, beginning at about 13:00 UT. Here, we use observations from the SECCHI/EUVI telescopes and other solar observations to analyze the behavior and geometry of the filament before and during the eruption. At this time, STEREO A and B were separated by about 8.5°, sufficient to determine the three-dimensional structure of the filament using stereoscopy. The filament could be followed in SECCHI/EUVI 304 Å stereoscopic data from about 12 hours before to about 2 hours after the eruption, allowing us to determine the 3D trajectory of the erupting filament. From the 3D reconstructions of the filament and the chromospheric ribbons in the early stage of the eruption, simultaneous heating of both the rising filamentary material and the chromosphere directly below is observed, consistent with an eruption resulting from magnetic reconnection below the filament. Comparisons of the filament during eruption in 304 Å and Hα show that when it becomes emissive in He II, it tends to disappear in Hα, indicating that the disappearance probably results from heating or motion, not loss, of filamentary material.

[1] Jet Propulsion Laboratory, California Institute of Technology, Pasadena, CA 91109, USA (email: Paulett.Liewer@jpl.nasa.gov)
[2] Naval Research Laboratory, Washington, DC 20375, USA
[3] Adnet Systems, Inc., Lanham, MD 20706, USA
[4] University College London, Mullard Space Science Laboratory, Holmbury St. Mary, Dorking, Surrey, RH5 6NT, UK
[5] Observatoire de Paris, LESIA, FRE 2461(CNRS), F-92195 Meudon Principal Cedex, France
[6] Konkoly Observatory of Hungarian Academy of Sciences, Budapest, Hungary

# 1. Introduction

The eruptions of filaments (also called prominences) have long been associated with the initiation of coronal mass ejections (CMEs) (Wilson and Hildner, 1986; Bothmer and Schwenn, 1994). Because of this close association with CMEs and the importance of CMEs in space weather, understanding the processes involved in filament eruptions continues to be an active area of research (see, for example, Tonooka *et al.*, 2000; Moore, Sterling, and Suess, 2007; Sterling *et al.*, 2007; Lites, 2008; Lin *et al.*, 2008); a better understanding of prominence eruptions may well lead to a better ability to forecast CMEs and associated space weather. The launch of the twin *Solar TERrestrial Relations Observatory* (STEREO) spacecraft in October 2006 has provided the opportunity to view a filament eruption from two viewpoints, giving new insights into the three-dimensional geometry. Here we use stereoscopic analysis of simultaneous extreme ultraviolet images from the STEREO A (Ahead) and B (Behind) spacecraft, in conjunction with other solar observations, to analyze a filament eruption that occurred on 19 May 2007 in an active region near disk center. This eruption was associated with a B9.5 flare, coronal dimming, an EUVI wave, and a double CME (Li *et al.*, 2008; Kilpua *et al.*, 2009; Veronig, Temmer, and Vrsnak, 2008). The evolution of the filament prior to the eruption has been studied by Bone *et al.* (2009).

Each spacecraft of the STEREO mission carries four remote sensing and *in situ* instrument suites (Kaiser, 2005). The *Sun Earth Connection Coronal and Heliospheric Investigation* (SECCHI) imaging package on each spacecraft includes six telescopes: an Extreme UltraViolet Imager (EUVI), inner (COR1) and outer (COR2) coronagraphs, and inner (HI1) and outer (HI2) heliospheric imagers (Howard *et al.*, 2008). The EUVI telescope has four channels similar to those of SOHO/EIT. The wavelengths and coronal temperature of peak response are 304 Å ($6-8 \times 10^5$ K, primarily the He II line), 171 Å ($10^6$ K, primarily Fe IX/X), 195 Å ($1.4 \times 10^6$ K, primarily the Fe XII line) and 284 Å ($2.2 \times 10^6$ K) (Wuelser *et al.*, 2004). The standard SECCHI synoptic program running during the eruption provided simultaneous A-B pairs at a ten minute cadence for 304 Å and 195 Å and a 2.5 minute cadence for 171 Å.

SECCHI data is available through the STEREO Science Center (http://stereo-ssc.nascom.nasa.gov).

Filaments lie above magnetic neutral lines separating large scale solar magnetic fields of opposite polarity and are observed in conjunction with very highly sheared magnetic field lines crossing the neutral line (Martin, 1998). They generally appear to consist of many "threads;" some threads appear to extend the full length of the filament, while others terminate elsewhere near the neutral line, giving rise to "barbs" (Martin, 1998). Filament plasma is denser and cooler ($10^4$ K) than the surrounding coronal plasma ($10^6$ K) and, on the disk, filaments generally appear in absorption in both H$\alpha$ and the extreme ultraviolet (EUV) lines. During episodes of filament "activation," however, filament threads are often seen in emission in various EUV lines, indicating that the material has been heated to coronal temperatures. Both ground and space observations, most recently from *Hinode* (see *e.g.,* Berger *et al.,* 2008*)*, have shown that the material is in constant motion in the threads, presumably moving along magnetic field lines. The volume filled with the highly sheared field lines running nearly parallel to the neutral line is generally referred to as the filament channel; the filament channel may exist with or without filament material (Martin, 1998). An arcade of coronal loops arches over the sheared "core" fields of the filament channel, connecting the opposite polarity regions.

In this paper, results are analyzed in the context of the standard model of filament eruptions with the goal of better understanding the physical processes occurring during the eruption and testing predictions from this model. In the standard model, a filament eruption is one part of a more general "unified" solar eruption that links filament eruptions, flares, CMEs, and post-eruptive arcades together as different manifestations of a single eruptive phenomenon driven by the buildup and release of magnetic energy by reconnection (see Shibata, 1999; Forbes, 2000; Priest and Forbes, 2002; Moore, Sterling, and Suess, 2007 and references therein). In this model, an eruption begins when the highly sheared non-potential magnetic field lines crossing the neutral line begin to reconnect with each other below the filament, presumably due to either increased stress in the magnetic field from emerging flux or chromospheric motions or decreased confining magnetic forces from the overlying arcade (from tether cutting or reconnection high in the corona). It should be noted that the scenario is the same whether or not filament material is actually present in the channel.. The reconnection between two highly sheared field lines crossing the neutral line - but with footpoints displaced along the neutral line - leads to one shorter field line crossing the neutral line and one longer loop more nearly aligned with the neutral line, which can rise somewhat due to its disconnection (see, *e.g.*, Figure 1 of Moore, Sterling, and Suess, 2007). A current sheet forms below the filament; reconnection in this current sheet leads to heating and acceleration of plasma, which often produces an X-ray flare. Some plasma is accelerated downward from the reconnection site, heating the chromosphere, leading to the formation of flare ribbons on either side of the neutral line and chromospheric evaporation. The reconnected flux below the filament is visible as hot (>1MK) post-eruptive arcades, which are seen as nearly-potential magnetic loops spanning the filament channel. These loops are filled by chromospheric evaporation. The reconnected flux above the reconnection point is locally detached from the surface in a full ejective eruption and begins moving outward, carrying some of the filamentary material along. The results presented in this paper are consistent with this scenario. Note that the standard model does not encompass an explanation for the underlying cause or "trigger" of the eruption. The underlying cause for the 10 May 2007 filament eruption has been discussed by Li *et al.*, (2008).

The STEREO/EUVI observations of the 19 May 2007 filament eruption allow us to use triangulation to track the height of the prominence during the eruption even though it occurs at disk center, thus enabling comparison with Hα disk observations as well. No dramatic rise is seen in the filament prior to the eruption, although during periods of "activation" prior to the eruption, a slight increase in the filament height is observed. Just prior to and at the start of the eruption, some of the filament threads are seen in emission in EUVI 304 Å; this hot filament material can be tracked as the one end of the filament rises to 1.08 $R_{sun}$ when the eruption begins. This heating is expected from the standard model since reconnection below the filament should lead to heating and accelerating plasma above the reconnection point as well as in the chromosphere below. The stereoscopic analysis shows that the prominence is still at approximately the same height when it begins to disappear in the Hα images, indicating that the disappearance of the filament in Hα is not due to depletion but rather due either to a heating of the material or mass motion shifting the Hα line out of the filter band pass. Dark ejected filamentary material seen in EUVI 304 Å can subsequently be tracked and triangulated to nearly 2 $R_{sun}$ about two hours after the start of the eruption, well after the filament has disappeared in Hα. Heating of the chromosphere extending much farther along the filament channel than seen in Hα is also observed; stereoscopic analysis allows us to confirm that the chromospheric ribbons lie directly below the location of the pre-eruption streamer along the neutral line. The evolution of the filament prior to the eruption is discussed in Bone *et al.* (2009) who report several filament-heating episodes seen in *Hinode*/XRT and STEREO/EUVI 171 Å.

In this paper, the stereoscopic tiepointing and triangulation technique used to analyze the STEREO/EUVI images is described in Section 2. Section 3 presents the results and analysis of three-dimensional (3D) reconstruction of the filament before, during, and after the eruption and compares the results with Hα observations from two ground-based observatories. Section 4 contains a discussion of the results.

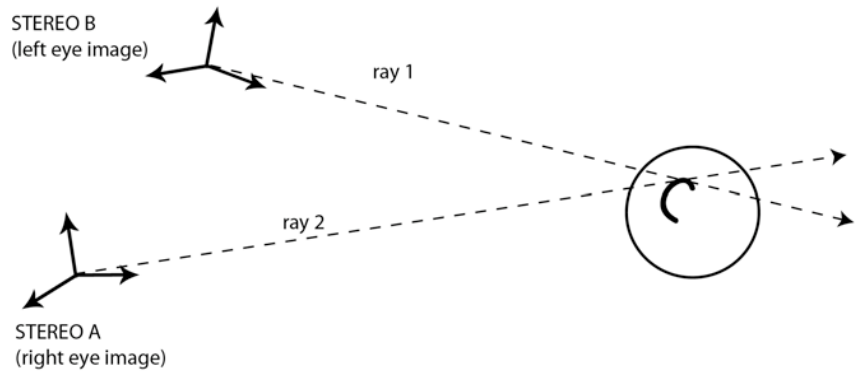

**Figure 1.** Schematic showing triangulation of a coronal feature from two spacecraft. Each pixel in a camera's image plane corresponds to a ray. If the same feature is identified in two images, the location of the feature can be determined from the intersection of the two rays.

## 2. Stereoscopic Data Analysis Technique

The two separated views of the solar corona provided by the twin STEREO A (Ahead) and B (Behind) spacecraft provide the opportunity to use stereoscopy for 3D reconstruction of solar features such as loops and filaments. If a feature can be identified in both images of a simultaneous pair, then the position of the object in a three-dimensional heliocentric coordinate system can be determined by triangulation provided that the location and separation of the two STEREO spacecraft are known, as well as the necessary information on the SECCHI cameras (plate scale and pointing relative to the Sun's center). Stereoscopy/triangulation is shown schematically in Figure 1. Each pixel on the telescope's image plane corresponds to a single ray, shown here intersecting a coronal loop. The feature recorded on A's image plane could lie anywhere along the ray and, similarly for the feature recorded on B's image plane. The rays intersect each other at the location of the feature and, thus, if the feature can be identified in both images of a simultaneous pair, its location in three dimensions can be determined by computing the intersection of the two rays.

The headers of the SECCHI image files contain all of the information on spacecraft location, pointing and plate scale (to convert pixel size to degrees) to the accuracy needed for 3D reconstruction (see Aschwanden *et al*., 2008). In performing the stereoscopic analysis, use is made of World Coordinate System (WCS) routines in the Solar Software Tree (also see Thompson, 2006), as well as the pointing and location information, to relate the coordinate system of the spacecraft's image plane to a heliocentric coordinates system and to "register" the images. Because the spacecraft are at different distances from the Sun, the images must also be scaled to make the pixel sizes identical. Results are generally given in either the Stonyhurst or Carrington heliocentric coordinate systems, but the WCS routines allow transformations to other systems as well. Stonyhurst coordinates are very similar to Carrington coordinates: radius and latitude are identical, but longitude is measured from the direction towards Earth in Stonyhurst (Thompson, 2006).

The 3D-reconstruction results presented in this paper were obtained by manually locating (and selecting with a cursor) the same feature in both images of a simultaneous pair; this is referred to as "tiepointing." Two different software tools for stereoscopy were used in results in this paper; one from the Solar Soft software library (scc_measure.pro, developed by W.T. Thompson) and one developed at JPL. Results from the two tools were compared to verify that both gave the same results for the 3D location of a solar feature. Both tools make use of an approximate "epipolar constraint" to aid in the placement of the tiepoints by reducing the placement from a 2D to a 1D problem: If the images were rotated so that a pixel row is parallel to the baseline between the two spacecraft, then the two tiepoints for a feature must lie in the same row (but different columns) in both images assuming affine geometry [see Inhester (2006) for a discussion of the epipolar constraint]. Thus, once a feature has been selected in one image, there is only one degree of freedom in placing the corresponding tiepoint in the second image. (Equivalently, the images are not rotated to the STEREO baseline, but after the feature is selected in the first image, the tool computes the approximate epipolar line and constrains the placement of the tiepoint in the second image to this line.) To reconstruct an extended object such as a prominence, a series of "tiepoints" along the object are identified and selected manually in each image. Each tiepoint pair is used to find the corresponding point in the 3D heliocentric coordinates system. The points are connected with straight-line segments to create the extended feature. No fitting or smoothing is used in the results presented here.

For a constant error in locating the tiepoints on the 2D images, the error in the 3D reconstruction decreases as the separation angle ($\varphi$) increases for a constant error. An error of one pixel in the placement of a tiepoint on the image leads to an error in the height of $\Delta h \approx \Delta x/\sin\varphi$ where $\Delta x$ is the error in placing the tiepoints. For SECCHI/EUVI images with a plate scale of $R_{sun} \approx 700$ pixels, an error of one pixel gives an approximate error of $\Delta h/R_S \approx 0.0014/\sin\varphi$. For May 2007, the separation between the A and B spacecraft was about 8°, leading to an error in the height determination of about $\Delta h/R_{sun} \approx 1\%$ if the feature can be tiepointed to within one pixel.

The limiting factor in reconstructing filaments from the EUVI data, however, was the ability to identify the same feature in both images of a STEREO pair. This can easily lead to an unknown error in excess of one pixel. Because much of the corona is optically thin, a bright (or dark) feature may be the result of a line-of-sight integration effect. Also, features may look different from different viewing angles, compounded by bright or absorbing foreground and background features along the ray paths (foreground/back-ground confusion). Dark filamentary material in 304 Å images can be difficult to distinguish from dark patches in the chromospheric network pattern. In many of the features reconstructed for this paper, it was not possible to reconstruct the entire length of a filament due to these effects. In particular, the filament's footpoints were often obscured by hot loops or lost in the chromospheric network pattern.

A sample stereoscopic analysis of the pre-eruption filament and the 3D reconstruction are shown in Figure 2. This analysis is for a pair of EUVI 304 Å images from 19 May 2007 at 11:24:15 UT, about one hour before the eruption. The left and center panels show the STEREO B (left eye) and A (right eye) images, respectively, both with the manually-placed tiepoints (white X's). The right panel shows two views of the 3D reconstruction of the filament; the straight lines in this and all images of the 3D reconstructions show the directions to STEREO A (red), Earth (green) and STEREO B (blue). The top shows the Earth view and the bottom shows an edge-on view. The highest part of the filament at this time is approximately at 1.06 $R_{sun}$ or 42 Mm above the solar surface.

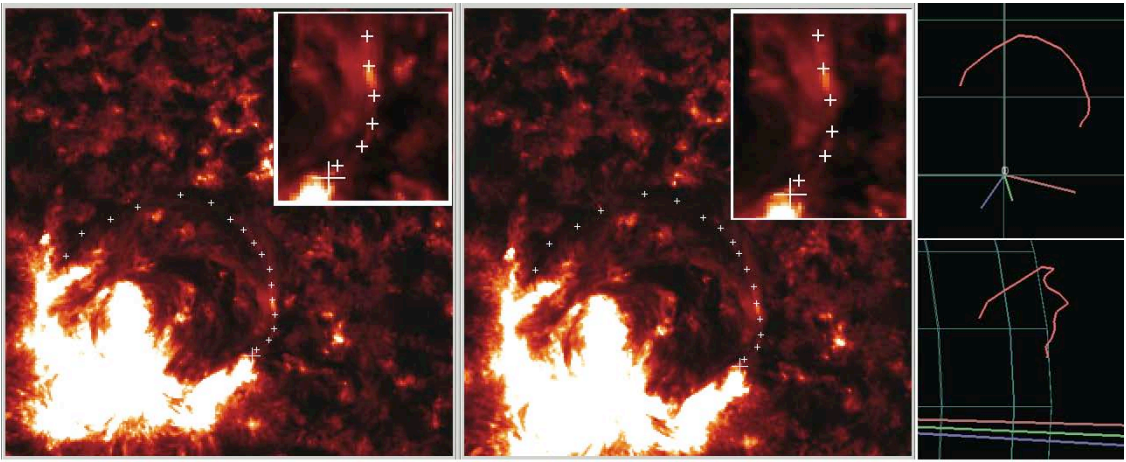

**Figure 2.** Tiepointing and reconstruction technique. The EUVI 304 Å images from Behind (left) and Ahead (center) for 11:24:15 UT on 19 May 2007 are shown with the tiepoints (white X's) used in the reconstruction. The insets are blowups of the set of lowest latitude points showing which features were tiepointed. The panel on the right shows the 3D reconstruction viewed from the front (top) and viewed edge-on (bottom). The three straight lines show directions to STEREO A (red), Earth (green) and STEREO B (blue).

# 3. Analysis of 19 May 2007 Filament Eruption

## 3.1 Context for the Eruption

On 19 May 2007 at approximately 12:50 UT, a filament eruption occurred in AR 10956 which was accompanied by a B9.5 flare, coronal dimming, and a double CME. A CME that has been associated with this eruption was seen by the *in-situ* instruments on STEREO B and Wind on May 22; STEREO A may have crossed the flanks of this CME (see Kilpua *et al.*, 2009). Thus this event is the first Sun-to-Earth solar event seen when the STEREO spacecraft had reached a significant enough separation to provide distinctly different views of the corona. The active region was near disk center. Thus stereoscopic analysis (triangulation) using STEREO observations allows us to study the true (3D) time-height evolution of this filament eruption and make comparisons with the evolution seen in Hα; a comparison not possible for limb prominence eruptions.

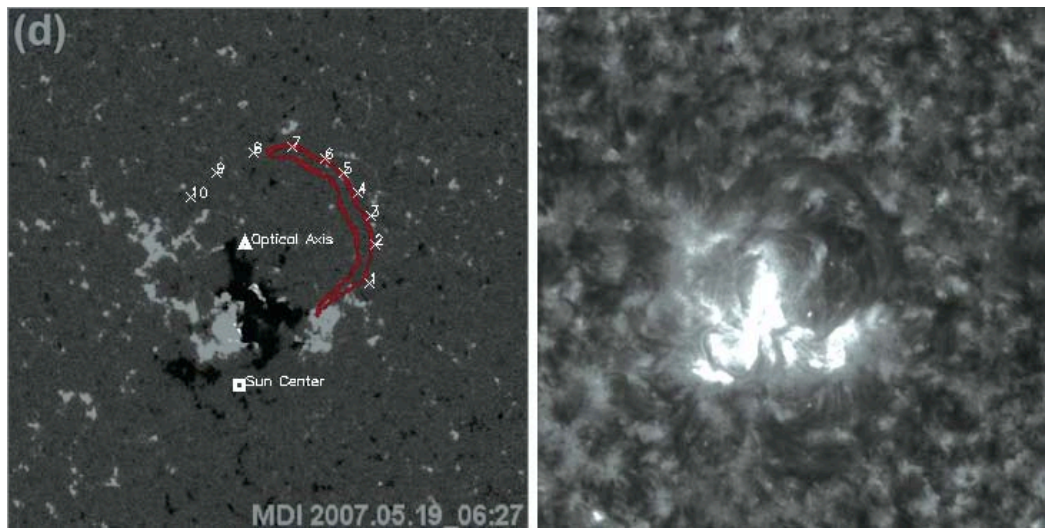

**Figure 3.** Left: Filament in relation to neutral line - MDI magnetogram at 06:27 with outline of Hα (red) from KSO 06:36 and tiepoints (white X's) used in the reconstruction of the filament from EUVI 304 Å images at 07:11:45 (based on Figure 4d from Li *et al.*, 2008). Right: STEREO A/ EUVI 304 Å at 07:11:45 showing the full extent of the filament.

Hα observations from the Kanzelhöhe Solar Observatory (KSO) show a continuous, large curved filament clearly at 12:32 UT with one end in AR 10956. The KSO Hα movie shows activity in the filament beginning just after this, leading up to the eruption. The filament begins to disappear in Hα at about 12:42 UT and flare ribbons begin to form at about 12:50 UT. The filament has essentially disappeared in Hα by 13:00 UT, although we are able to track it stereoscopically with STEREO EUVI 304 Å for about two more hours. The B9.5 GOES soft X-ray flare begins at 12:48 UT, as the flare ribbons begin to form, and peaks at 13:00 UT. The RHESSI hard X-ray flux starts rising at 12:50 UT and peaks at about 12:52 UT. The corresponding STEREO observations of the filament will be discussed below in Section 3.2. After the eruption, STEREO EUVI 195 Å observations show an EUV wave (Veronig, Temmer, and Vrsnak, 2008), coronal dimming and post-eruptive arcade formation at the neutral line. The coronal context of this eruption has been described in detail in Li *et al.* (2008). The left side of Figure 3, based on Figure 4d from the Li et al. paper, shows the MDI magnetogram at 06:27 UT on 19 May overlaid with the Hα filament (red outline – from the KSO Hα image) and the tiepoints (white X's) used in our 3D reconstruction of the filament at 07:11 UT on 19 May from the STEREO/EUVI 304 Å images. On the right is the EUVI 304 Å image at 07:11 UT from STEREO A. It can be seen that the filament begins in the active region then curves around to the Northeast into a weak-field region following the neutral line. The filament can be seen to extend further in the EUVI 304 Å images than in the Hα image.

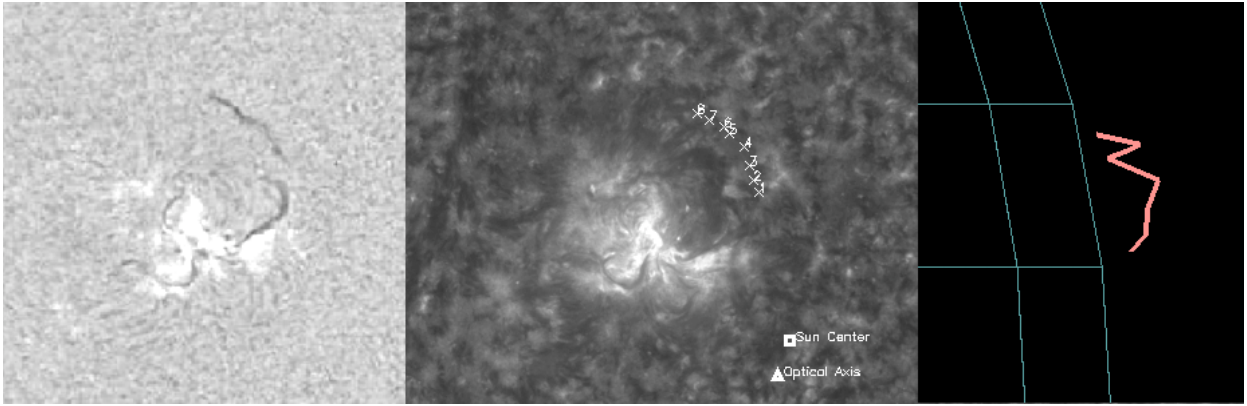

**Figure 4.** Left: Hα observation from KSO showing two separate filaments on 18 May at 10:36. Center: STEREO B/ EUVI 304 Å image at 10:35:45 with tiepoints used in the reconstruction of the upper filament. Right: 3D reconstruction of the upper filament as viewed from the side.

## 3.2 Stereoscopic Analysis of the Filament

The day prior to the eruption, 18 May, Hα observations from KSO show two separate filament segments at 10:36 UT (Figure 4, left panel). The center panel shows the STEREO B/ EUVI 304 Å at 10:35:45 UT with the tiepoints used in the reconstruction. The right panel shows the 3D reconstruction at this time as viewed from the side. At this time, only the upper filament could be seen well enough in EUVI 304 Å to perform a 3D reconstruction. By 11:30 UT, the SECCHI/EUVI 304 Å observations suggest that gap between the filaments has been bridged; at this time EUVI 304 Å (see supplemental movie 1) show a flow of material moving along filament threads across the gap seen in the earlier Hα image at 10:36 UT (Figure 4). By 11:00 UT on 18 May, we were able to reconstruct the longer "merged" filament using simultaneous EUVI 304 Å pairs and the technique described in Section 2. The merging of the two filaments has been analyzed in detail by Bone *et al.* (2009).

The filament was reconstructed via the tiepointing technique for over 30 different times on 18 and 19 May to study the evolution, primarily using EUVI 304 Å images pairs. Generally, the filament is seen in EUVI 304 Å, responsive to a temperature range $6 - 8 \times 10^5$ K, as a dark absorbing feature consisting of multiple threads (*cf.* Figure 2). Analysis of the 3D reconstructions on the dark filamentary feature from 11:30:45 UT on 18 May to just before the eruption on 19 May indicate that the dark feature seen in EUVI 304 Å changed little in overall geometry during this time, maintaining a similar shape and height throughout within the uncertainties of the tiepointing process (several pixels due to the width of the filament as well as foreground/background confusion). This is in contrast to the filament's appearance in Hα observations from the Mauna Loa Solar Observatory (http://mlso.hao.ucar.edu/cgi-bin/mlso_data.cgi), which show considerable variation on 18 May from 16:30 - 21:24 UT. Generally, during this time, the MLSO filament shows two or more "segments," but sometimes these "join" to form a continuous structure as seen in the EUVI 304 Å images. The behavior of the filament during this period is analyzed using additional data from *Hinode* and other ground-based Hα observatories in Bone *et al.* (2009).

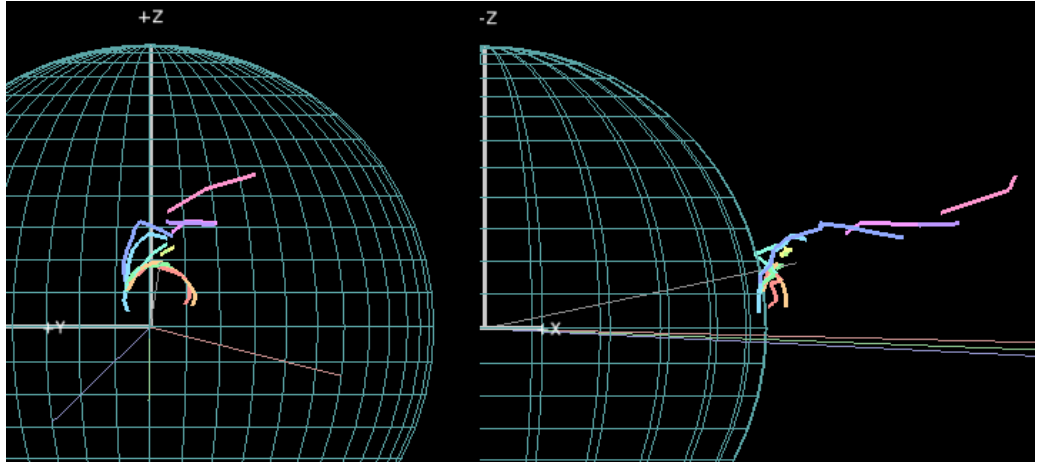

**Figure 5.** Two views of 3D reconstructions of the filament at 11 times during the eruption; each colored segment is the reconstruction at a different time. It can be seen that the entire length of the filament erupts with the southwest end in the active region rising faster. Some counter-clockwise rotation of the filament is evident. (The 3D object graphic is shown in rotation in the supplementary material, movie 2).

Reconstructions from 11 times on 19 May (from 12:41:45 UT to 14:41:45 UT) are shown in Figure 5 where each color indicates a 3D reconstruction at one particular time. The left and right panels show this set of 3D reconstructions from two different viewpoints. (Supplementary movie 2 is an animation showing a rotating view of these 3D reconstructions.) The eruption is clearly very asymmetrical with the southwest end of the filament in the active region rising faster; the weak-field end appears to remain attached to the chromosphere. The flare ribbons appear first in the active region, also indicating that the eruption begins here. The asymmetry in the eruption leads to a downflow of filamentary material towards the northeastern end of the channel. The 3D reconstructions shown in Figure 5, especially those at 13:11:45 UT (blue) and 13:21:45 UT (purple) show that a single very asymmetric whip-like filament eruption occurs. The reconstruction at 13:11:45 UT is also shown in Figure 11 and discussed in Section 3.3.

Bright chromospheric ribbons seen in all EUVI channels continue to form along the filament channel, progressing northeasterly along the channel as the asymmetric eruption progresses from strong to weak field regions. The maximum extent of the ribbons is seen in EUVI 304 Å at 14:11:45 UT; the ribbon is quite patchy at this time. Comparison of the 3D reconstructions of these ribbons (lying at $\approx 1 R_{sun}$) with the 3D reconstructions of the filament prior to eruption allows us to confirm that the chromospheric ribbons lie at low altitudes below the location of the pre-eruption streamer along the neutral line. This is consistent with the standard model of a filament eruption in which magnetic reconnection occurs in highly sheared field lines below the filament. Plasma is heated and accelerated at the reconnection site, some of which flows downward, causing the flare, heating the chromosphere and giving rise to the bright ribbons. These ribbons mark the footpoints of reconnected magnetic field on either side of the filament channel. The reconnected magnetic flux is best visible in the EUVI 195 Å images as a post-eruptive arcade of loops spanning the channel at the former location of the filament (see Figure 3 in Kilpua *et al*., 2009). Above the reconnection site, the magnetic field threading the filament begins to move upward, carrying the filamentary material with it. EUVI observations during the eruption show that some of the rising filamentary material is now seen in emission, indicating that it has also been heated (discussed further in Section 3.3). This is also consistent with the standard model since reconnection below the filament will also accelerate plasma upward, heating the trapped material. Further evidence for reconnection below the filament comes from stereoscopic reconstruction of the early post eruptive loops seen in EUVI 171Å (at 13:11:30 UT) and 195 Å (at 13:12:00 UT and 13:42:00 UT). These loops are between 10 - 20 Mm, which is less than the typical height of the prominence before the eruption of 28 - 50 Mm. While this is consistent with reconnection below the filament, it is not definitive since the reconnected flux loops may have shrunk.

Figure 6 summarizes the results of 20 3D reconstructions on 19 May in a height – time plot of the 3D trajectory of the filamentary material. Between 12:00 UT and 14:00 UT, all simultaneous EUVI 304 Å STEREO pairs were analyzed; the cadence at this time was ten minutes. Here we plot the highest (largest radial distance) reconstructed point in each 3D reconstruction file and thus it does not necessarily represent a tracking of the same material. Moreover, this point may not be the highest point on the filament because not all portions of the filament could be tiepointed at each time due to foreground and/or background confusion. There is some evidence in Figure 6 for a slight, slow rise before the main eruption. Similar changes in height were seen earlier during periods of surge-related activation, as discussed below, and thus this does not always indicate that eruption is imminent. The rapid rise from one to two solar radii immediately follows the flare peak at 13:00 UT. Arrows show the time of the soft X-ray flare rise (12:48 UT) and peak (13:00 UT).

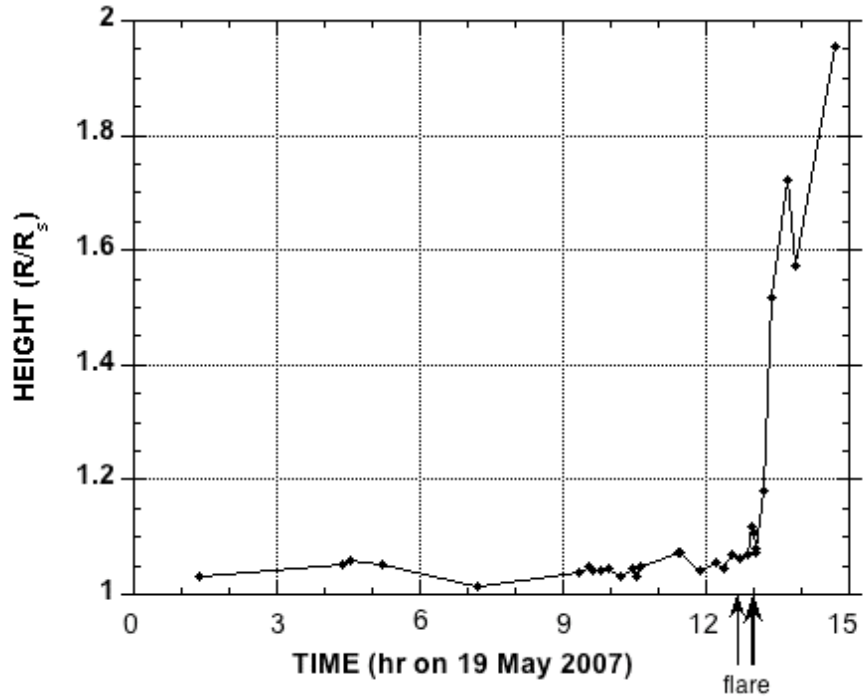

**Figure 6.** Trajectory of the filament as determined from the 3D reconstructions. The small (large) arrow marks the start (peak) of the B9.5 flare. The slight rises in height prior to eruption (≈04:30 and ≈11:30) occur during periods of surge-like activation.

Filamentary material could be tiepointed and reconstructed only until 14:41:45 UT. After this time, background confusion made it impossible to tiepoint the material, although it could still be seen in movies of the data until 15:51:45 UT. This is clear in both monoscopic and anaglyph EUVI 304 Å movie (needs red/blue glasses; see supplemental movies 3 and 4 respectively). Two CMEs in the LASCO CME catalog (http://cdaw.gsfc.nasa.gov/CME_list/) occurred at the same time as this filament eruption and flare; these CMEs have been interpreted as a double CME originating in association with this eruptive event (Kilpua *et al.*, 2009). Although the events are happening at the same time, the filamentary material lags far behind the CME fronts tracked in LASCO. The relationship between the X-ray flux histories and the CME fronts can be seen in Figure 6 of Veronig, Temmer, and Vrsnak (2008). The radial propagation vector of the filamentary material, determined from analysis of the reconstructions from 12:56 to 14:41 UT was 13° latitude and -1° longitude in heliographic

Stonyhurst coordinates (Earth at 0° longitude). The average velocity of 103 km s$^{-1}$ is well below the speed of the associated CMEs, but if the fast associated CME (≈900 km s-1) were also propagating in this direction, it is consistent with this CME reaching STEREO B and *Wind* 2.7 days later as observed (Kilpua *et al.*, 2009).

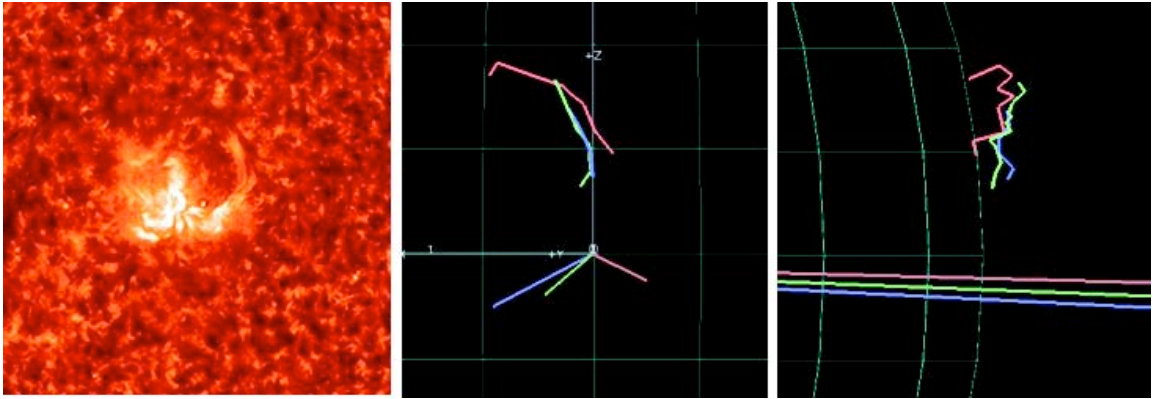

**Figure 7**. Left: Filament in EUVI 304 Å image during a heating episode at ≈4:30 UT on 19 May 2007 showing both absorbing and emissive threads. Center and Right: Two views of 3D reconstructions comparing heated threads at 04:32:00 UT (195 Å -blue) and 04:31:45 UT (304 Å -green) with dark filament reconstructed from 304 Å images at 04:21:45 UT (red, ten minutes earlier). The heated threads seen in emission are somewhat higher than the dark absorbing filament.

The EUVI 304 Å images show that several episodes of filament activity occurred during the period 24 hours preceding the eruptions, when hotter material, seen in emission, appears to surge up along filament threads. The emissive threads can also be seen in both EUVI 195 Å and 171 Å (≈1 MK), indicating that some of the material has been heated to coronal temperatures. One such activation also immediately preceded the eruption. In all cases, the hotter material seemed to originate from the end of the filament rooted in the active region. Such surge-like activations of filaments unrelated to eruptions are common and frequently cause the filament to extend to greater height than before the surge (see review by Martin, 1989). These surges may result from reconnection at or below the base of the corona (Lin *et al.*, 2008). Figure 7 shows results from a stereoscopic analysis of a surge-related heating episode that occurred at approximately 04:30 UT, eight hours prior to the eruption. The panel on the left shows the filament 04:31:45 UT in EUVI 304 Å; both dark absorbing and bright emissive threads can be seen. The panels on the right show two views of the 3D reconstructions of the dark filament at 04:21:45 UT (red, prior to the heating) and the hot emissive filament at 04:31:45 UT (green, reconstructed from EUVI 304 Å) and at 04:32:00 UT (blue, reconstructed from EUVI 195 Å). [In the reconstructions in Figures 7 - 10, the three thin lines starting from the origin indicate the directions to STEREO B (purple), Earth (yellow-green) and STEREO A (red).] The emission features seen in He II and Fe XII apparently correspond to the same thread; this thread is close to, but perhaps slightly higher than, the dark threads, consistent with observations reviewed in Martin (1989). A similar heating episode occurred at about 11:00 UT; a bright thread reconstructed at 11:26 UT can be seen as a somewhat higher point (1.07 $R_{Sun}$) in Figure 6. The final heating episode leading up to the eruption began at about 12:30 UT.

Throughout this paper, we have made the assumption that features seen in emission in 304Å (primarily the He II line) have been heated to about 6-8 x $10^5$ K, which is the case for a collisional equilbrium situation. In fact, the situation is more complex (see Labrosse, Goutebroze, and Vial, 2007) in a non-equilibrium prominence with rapidly moving material. Addressing the details of line formation is beyond the scope of this work. However, the presence of He II in the filament both in the early interactions and in the eruption itself is strongly suggestive of energy release due to magnetic reconnection. Moreover, emission is generally seen simultaneously in the other "hotter" EUV channels, and thus it is reasonable to interpret the He II emission as an indication of heating .

### 3.3 Comparison of Stereoscopic Observations with Hα Observations

In this section, we compare STEREO/EUVI 304 Å, 171Å, and KSO Hα observations to understand how the evolution in Hα relates to the processes as observed in EUVI. The comparison is limited by the low ten min cadence of EUVI 304 Å. Here we compare observations taken within one second of each other.

In Figure 8, the filament at 12:32 UT is compared in Hα (top left), and EUVI 304 Å (center) and 171 Å (top right) images. The two lower panels show 3D reconstructions from EUVI 304 Å (red) and 171 Å (blue) images from viewed from Earth (left) and edge on (right). The full extent of the filament is seen better in He II than in Hα, as was the case on 18 May as well (Section 3.2). The portion not visible in Hα is fainter in EUVI 304 Å image (Figure 8, top row) and the 3D reconstructions from the EUVI 304 Å images indicates that this portion is also somewhat lower (height of ≈1.02 $R_{sun}$ *versus* heights of 1.05 - 1.1 $R_{sun}$). The 3D reconstruction of the bright thread seen in 171 Å images coincides with the 3D reconstruction from 304 Å images.

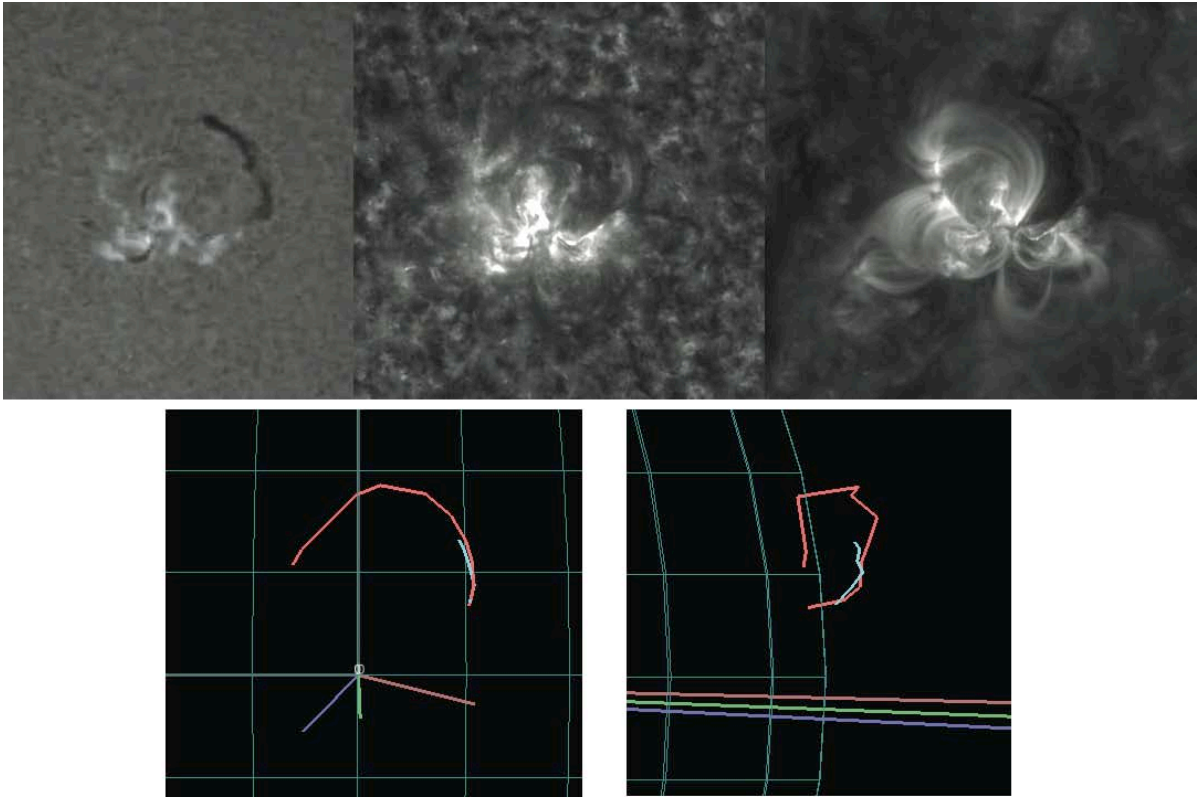

**Figure 8.** Top: Comparison of filament as seen in Hα (left), EUVI 304 Å (middle) and 171 Å (right) images at 12:32 UT on 19 May 2007; the full extent of the filament is best seen in EUVI 304 Å image. Bottom: The 3D reconstructions show that the northeastern portion of the filament not seen in Hα is lower than the other end.

A comparison ten minutes later, at 12:42UT, is shown in Figure 9, with the Hα image on the left and the EUVI 304 Å image on the right. The boxed region in the 304 Å image shows a number of hot bright emissive threads now in the region of the filament near the southwest end in the active region. In Hα, this portion has become faint and diffuse and suggests that the disappearance in Hα is not due to depletion, but rather due either to a heating of the material or mass motion shifting the Hα line out of the filter band pass. This behavior has been observed several times previously in this region by Bone *et al*. (2009). However the occurrence indicated in Figure 9 marks the "onset" phase of the filament eruption (Moore, Sterling and Suess, 2007), when reconnection is just beginning below the filament in the strong field region, leading to some heating of filamentary material above the reconnection site. These heating events are probably caused by flux cancellation occurrences.

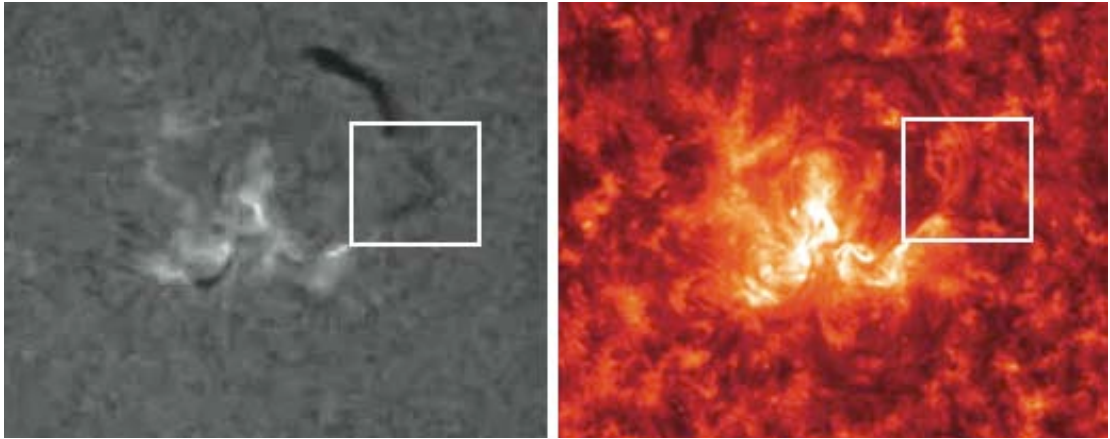

**Figure 9.** Comparison of the filament as seen in Hα (left) and EUVI 304 V (right) at 12:42 UT on 19 May 2007. Bright emissive threads in EUVI 304 Å indicate that the fading of the filament in Hα is due to heating (or motion), not depletion.

At 12:52 UT (Figure 10), the eruption has begun. Flare ribbons are clearly visible as bright features in Hα (top left, white arrows) as well as in the EUVI 304 Å (center) and 171 Å (right) images. While only a small patch of the filament is visible in Hα, we are able to see and reconstruct a long segment of the erupting filament in 3D from EUVI 304 Å images (lower panels, purple segment). This 3D reconstruction shows that the filament is erupting along most of its length. The end of the filament in the strong-field region has now risen to about 1.06 $R_{sun}$, while the lowest point tiepointed in the weak-field region was at about 1.005 $R_{sun}$, again showing the asymmetry of the eruption. The two lower panels in Figure 10 show 3D reconstructions of other features visible in the EUVI 171 Å data. The red and blue segments are 3D reconstructions of the chromospheric flare ribbons (height of 1.0 $R_{sun}$, as expected). The green segments represent the 3D reconstruction of the bright "hot spots" seen in 171 Å at a height of about 1.07 $R_{sun}$, coinciding in 3D with the filament reconstructed from EUVI 304 Å (purple segment). This shows that there is hot material moving outward with the rising magnetic flux. Thus, STEREO/EUVI observations show both aspects of the reconnection heating expected from the standard model: Not only has reconnection led to the heating associated with the flare ribbons (at 1 $R_{sun}$), but has also led to heating of filamentary material threaded by the reconnected magnetic field moving upward (up to 1.07 $R_{sun}$). Only because of the stereoscopic observations provided by STEREO/EUVI are we able to determine that some bright emissive features seen in He II are in the chromosphere and others in the corona. Again, the disappearance in Hα appears to be caused by heating or motion of the filamentary material rather than depletion.

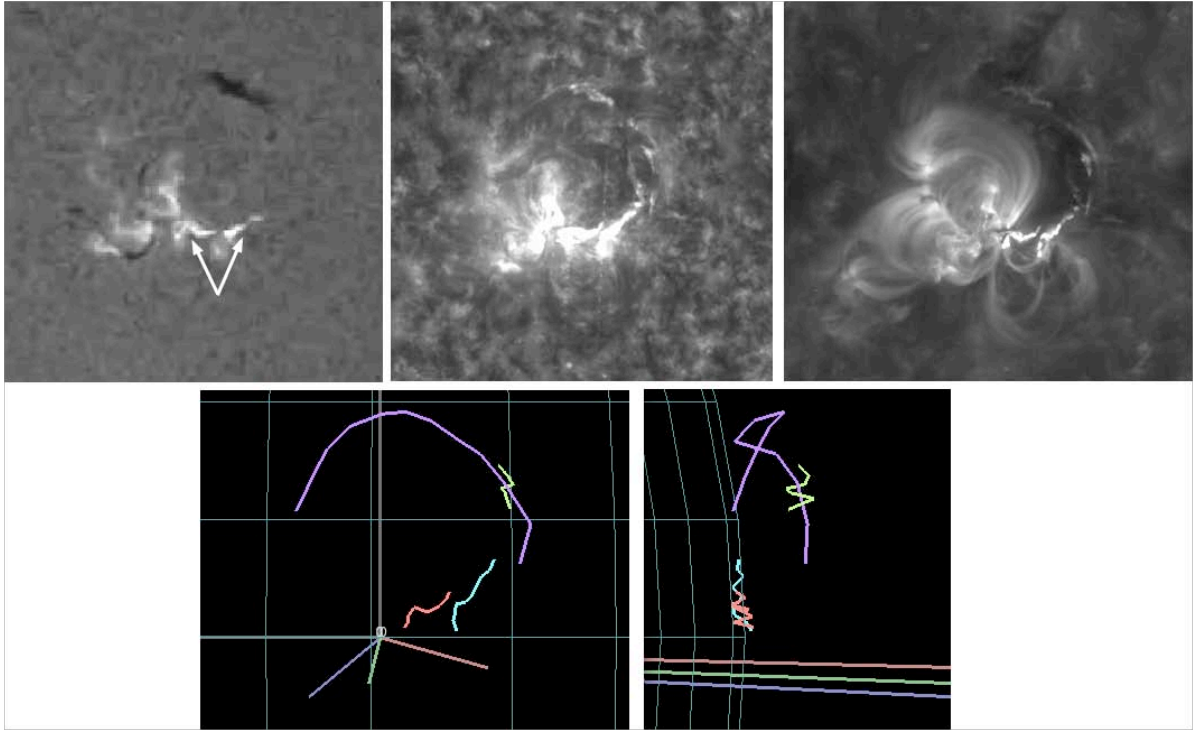

**Figure 10.** Top: Hα (left), EUVI 304 Å (center), and 171 Å (right) images at 12:52 UT. Both EUV images show hot filamentary material in the erupting filament not seen in the Hα image. Bottom: Two views of the 3D reconstructions of the erupting filament from EUVI 304 Å (purple), "hot spots" from EUVI 171 Å and the chromospheric ribbons (red and turquoise) from EUVI 304 Å. The filament erupts along almost the entire length with the active region end rising faster.

The final comparison is made after the eruption at 13:12 UT when the filament has almost completely disappeared in Hα (Figure 11, top left). A long segment of the erupting filament can be seen as an absorption in the EUVI 304 Å image (arrows from top – better seen in supplemental movies 3 and 4); the two lower panels show two views of the 3D reconstruction of the erupting filament (green segment). This also shows that the filament has erupted along much of its length in a single eruptive event. The highest point in this reconstruction is now at about 1.18 $R_{sun}$. The lower portion, best seen in the edge on view, still lies at about 1 $R_{sun}$, suggesting that this end has perhaps remained tied in the chromosphere. The red line segment in the two lower panels is a reconstruction of the bright east flare ribbon (at ≈1 $R_{sun}$) visible in both He II (arrow from right) and Fe IX/X, which extends farther than the flare ribbons seen in Hα (top left), indicating that reconnection has progressed farther along the filament channel than suggested by the Hα ribbons. (Note, however, that reconnection in weak-field regions may not produce enough heating to lead to chromospheric ribbon formation.) At this time as well, stereoscopic analysis has allowed us to determine that the dark feature (the erupting filament) is at coronal heights whereas the bright feature (the patchy ribbon) lies in the chromosphere.

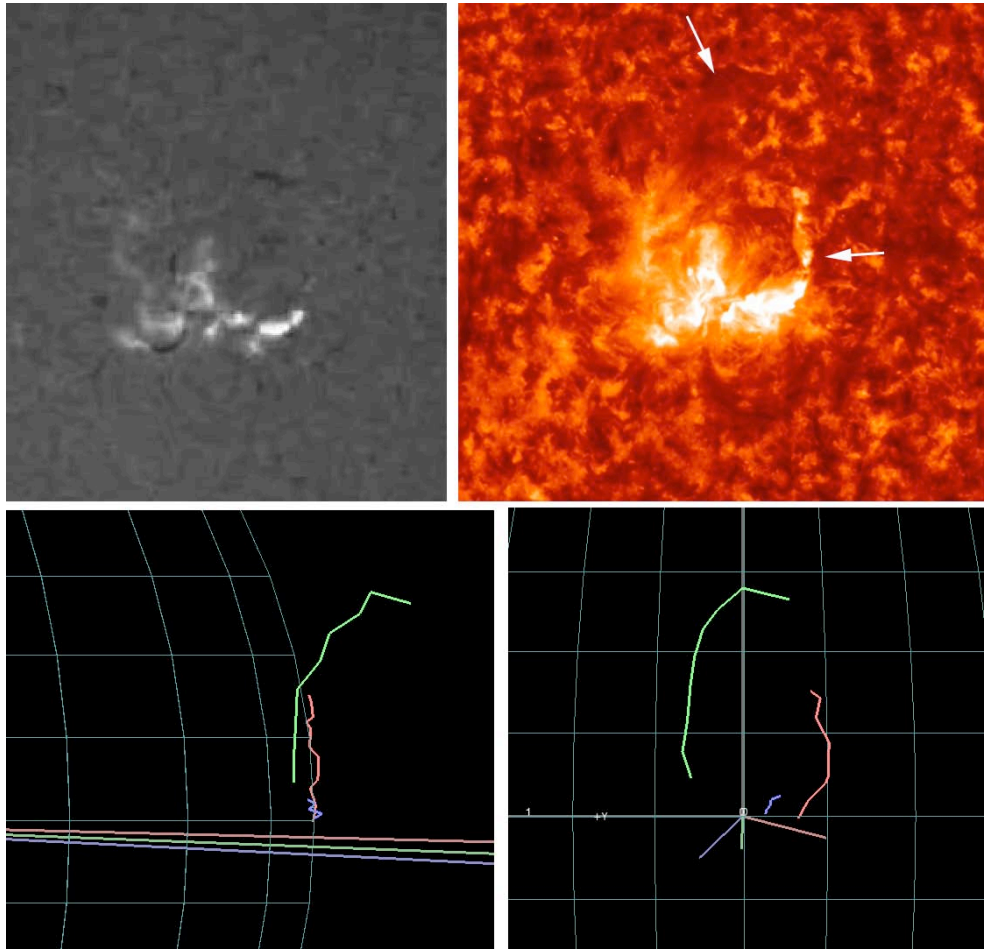

**Figure 11.** Top**:** Hα (left) and EUVI 304 Å (red) images at 13:12 UT. The filament has almost completely disappeared in Hα; it is visible in the EUVI 304 Å image only in absorption (top arrow). Bottom: The two panels show 3D EUVI 304 Å reconstructions from two viewpoints of the erupting filament (green) and the chromospheric ribbons (blue and red).

## 4. Discussion

We have used stereoscopic analysis of STEREO/EUVI data in conjunction with Hα data to study the evolution of the prominence in AR 10956 which erupted at about 12:52 UT on 19 May 2007. This eruption was associated with a B9.2 flare, a post-eruptive flare arcade, dimming, an EUV wave and a double CME and related ICMEs/MCs (see Veronig, Temmer, and Vrsnak, 2008; Li *et al.*, 2008; Kilpua *et al.*, 2009). 3D reconstructions of the erupting filament indicate that the filament erupted in a single event in an asymmetric whip-like fashion with the end in the active region rising fastest and the weak field end perhaps remaining rooted in the chromosphere. Stereoscopic analysis allowed us to follow the trajectory of the erupting filament in three-dimensions, determining both the speed and direction, up to a height of 1.08 $R_{sun}$. The speed (≈100 km s-1) was well below that of the associated CMEs.

Gissot et al. (2008) have also analyzed this filament eruption using an entirely different stereoscopic technique for 3D reconstruction. Their techniques used an optical flow method to automatically find displacements between features in two images of a stereoscopic pair using SECCHI/EUVI 304 Å images. Comparison with their results for the highest point on the filament at 12:41:45 UT shows that the results from the two techniques are in good agreement (~44Mm).

Results of the stereoscopic analysis can be used to test several predictions of a unified solar eruption model (*e.g.* Forbes 2000), which relates flares, filament eruptions, and CMEs. In this model, magnetic reconnection *below* the filament causes the ejective eruption, often causing a flare. Reconnection below the filament should lead to reconnection flows, and thus plasma heating, both above and below the reconnection site. STEREO/EUVI 304 Å observations show emissive feature indicating heating both of the chromosphere below (the chromospheric "flare" ribbons) and of the filamentary material above, which is rising during the eruption. Only with STEREO can we verify that some emissive features (the ribbons) are at chromospheric heights while others, seen at the same time, are at coronal heights (the rising filament). Further evidence suggesting reconnection below the filament comes from 3D reconstructions of early post-eruptive arcade loops whose heights are found to be less than that of the pre-eruption filament.

By comparing 3D reconstructions of the filament prior to the eruption with the chromospheric ribbons after the eruption, we verify that the post-eruptive ribbons are located directly beneath the former location of the filament (presumably on either side of the polarity inversion line) as predicted by the standard model. A patchy chromospheric ribbon seen in EUVI 304 Å images extends much further along the filament channel than the corresponding H$\alpha$ ribbons. Prior to eruption, the filament as seen in EUVI 304 Å images also extends much further along the neutral line than that seen in H$\alpha$. Comparison of EUVI 304 Å and H$\alpha$ images shows that, at times, disappearance of the filament in H$\alpha$ results from heating of the filamentary material (or motion out of the filter band pass) and not depletion.

The stereoscopic analysis has not shed light on the underlying cause of the eruption, which followed a surge-related heating of the filament. Several such episodes had been observed in the proceeding 12 hours, suggestive of the "confined" eruptions described by Moore, Sterling, and Suess, 2007. It is not uncommon to observe filament activation by surges not leading to eruptions (see review by Martin, 1989). Future observations from STEREO, both using triangulation or more generally, being able to see filaments and eruptions from two widely separated view points, may provide a better understanding of the underlying cause of solar eruptions.

*Acknowledgments*

We thank K.E.J. Kilpua, Y. Li, J. Luhmann, B. Lynch, S. Martin, O. Penasco, and A. Vourlidas for very valuable conversation on this solar event. We thank N. Rich, S. Suzuki, and J.-P. Wuesler for their help with the SECCHI/EUVI data. The *STEREO*/SECCHI data used here are produced by an international consortium of the Naval Research Laboratory (USA), Lockheed-Martin Solar and Astrophysics Lab (USA), NASA Goddard Space Flight Center (USA), Rutherford Appleton Laboratory (UK), University of Birmingham (UK), Max-Planck-Institut für Sonnensystemforschung (Germany), Centre Spatiale de Liege (Belgium), Institut d'Optique Théorique et Appliqueé (France), Institut d'Astrophysique Spatiale (France). The USA institutions were funded by NASA; the UK institutions by Particle Physics and Astronomy Research Council (PPARC); the German institutions by Deutsches Zentrum für Luft- und Raumfahrt e.V. (DLR); the Belgian institutions by Belgian Science Policy Office; the French institutions by Centre National d'Etudes Spatiales (CNES) and the Centre National de la Recherche Scientifique (CNRS). The NRL effort was also supported by the USAF Space Test Program and the Office of Naval Research. A portion of this work was carried out at the Jet Propulsion Laboratory, California Institute of Technology under a contract with NASA. JLC acknowledges the award of a Leverhulme Emeritus Fellowship.
institutions by Deutsches Zentrum für Luft- und Raumfahrt e.V. (DLR); the Belgian institutions by Belgian Science Policy Office; the French institutions by Centre National d'Etudes Spatiales (CNES) and the Centre National de la Recherche Scientifique (CNRS). The NRL effort was also supported by the USAF Space Test Program and the Office of Naval Research. A portion of this work was carried out at the Jet Propulsion Laboratory, California Institute of Technology under a contract with NASA. JLC acknowledges the award of a Leverhulme Emeritus Fellowship.